\documentclass[aip,preprint,letterpaper,groupedaddress,longbibliography]{revtex4-2}
\usepackage{amsmath,amsthm,amsfonts}
\theoremstyle{definition}

\usepackage{multirow}
\usepackage{array}
\usepackage{soul}
\usepackage{float}
\usepackage{graphicx}
\usepackage{afterpage}
\usepackage{tabularx}
\usepackage{siunitx}
\usepackage{amsmath}
\usepackage{textcomp}

\usepackage[font=small,labelsep=quad,format=plain]{caption}
\usepackage{subcaption}
\usepackage[ colorlinks=true,citecolor=blue,linkcolor=blue,urlcolor=blue]{hyperref}
\usepackage{cleveref}
\crefname{equation}{Eq.}{Eqs.}
\crefname{section}{Sec.}{Secs.}
\crefname{subsection}{Sec.}{Secs.}
\crefname{appendix}{Appendix}{Appendices}
\crefname{figure}{Fig.}{Figs.}
\crefname{table}{Table}{Tables}
\crefname{proposition}{}{}
\crefname{corollary}{}{}

\usepackage{bm}
\usepackage[dvipsnames]{xcolor} 

\begin{document}

\title{Thermal laser evaporation of superconducting Nb films}

\author{Yifei Yan}
\affiliation{Division of Engineering and Applied Science, California Institute of Technology, Pasadena, CA 91125, USA}
\author{David S. Catherall}
\affiliation{Division of Engineering and Applied Science, California Institute of Technology, Pasadena, CA 91125, USA}
\author{Finley B. Donachie}
\affiliation{Division of Engineering and Applied Science, California Institute of Technology, Pasadena, CA 91125, USA}
\author{Dorsa Najafipanah}
\affiliation{Division of Engineering and Applied Science, California Institute of Technology, Pasadena, CA 91125, USA}
\author{Austin J. Minnich}
\email{aminnich@caltech.edu}
\affiliation{Division of Engineering and Applied Science, California Institute of Technology, Pasadena, CA 91125, USA}
\date{\today}

\begin{abstract}
Thermal laser evaporation (TLE) is an emerging physical vapor deposition (PVD) technique that uses continuous-wave lasers to evaporate even the most refractory elements. The deposition and characterization of superconducting refractory elements are of interest to assess the quality of films deposited by TLE.  Here, we report the deposition of niobium (Nb) on c-plane sapphire using a home-built TLE system equipped with a 1070~nm continuous-wave fiber laser. A deposition rate of 0.19 \AA/s was achieved with 533 W of laser power focused to a diameter of approximately 500 \textmu m on a Nb target. A film with a thickness of $22.47 \pm 0.05$~nm and a root-mean-square (RMS) roughness of $0.23 \pm 0.02$~nm exhibited a room-temperature resistivity of $21.60 \pm 0.05$~\textmu$\Omega \mathrm{cm}$ and a residual-resistivity ratio (RRR) of 5.71. The superconducting transition temperature, $T_\mathrm{c}$, of the film was measured as 9.24~K, only 10 mK below the bulk value, with a transition width of 29 mK. Among films deposited without substrate heating, the electrical and superconducting properties of the present films meet or exceed those deposited by other techniques. This work contributes to the emerging consensus regarding TLE as a versatile PVD method for refractory elements and compounds.

\end{abstract}

\maketitle
\newpage

\section*{Introduction}

Niobium (Nb) is a refractory transition metal with the highest bulk superconducting transition temperature, $T_\mathrm{c}$, of 9.25~K among all  elemental superconductors \cite{waldram2017superconductivity}. Its low quasiparticle loss at accessible temperatures below $T_\mathrm{c}$ \cite{tinkham1996introduction} and well-established thin-film growth techniques \cite{fowler1963superconducting, mayadas1972electrical, wu1979intrinsic, face1983ion, anders2007high, Smart:2021} have made it an important material for superconducting electronics. In thin-film form, Nb is widely employed in superconducting microwave resonators \cite{wallraff2004strong}, Josephson junctions \cite{tolpygo2019advanced}, and transmon qubits\cite{anferov2024improved}, where device performance is sensitive to the superconducting properties and microstructure, as well as the quality of the surfaces and interfaces of the film \cite{altoe2022localization, de2021materials, oh2024exploring}.

Nb films are commonly prepared by sputtering and electron-beam evaporation, both of which have produced superconducting films suitable for device fabrication \cite{premkumar2021microscopic, kowsari2021fabrication}. In sputtering, plasma-generated energetic species and gas-phase collisions introduce additional kinetic parameters that influence film microstructure and residual stress \cite{windischmann1992intrinsic}. Electron-beam evaporation enables thermal evaporation of refractory materials but introduces additional process complexity. Electron-beam sources exhibit flux instabilities and therefore may require active monitoring and feedback control \cite{chambers2000epitaxial}. For refractory metals, achieving sufficiently high vapor fluxes for deposition requires high source temperatures and beam powers, necessitating the consideration of source containment, crucible compatibility, and contamination control \cite{mattox2010pvd}. In addition, secondary and backscattered electrons as well as ionized vapor species generated during evaporation can interfere with lithographic processing, causing deposition beneath resist undercuts, resist damage, and lift-off defects \cite{volmer2021solve}. 


Thermal laser evaporation (TLE) is an emerging physical vapor deposition (PVD) technique that uses continuous-wave lasers to heat a localized region of a source, producing vapor for deposition. The surrounding material outside of the irradiated region can remain solid, allowing many elements to be evaporated from free-standing sources without crucibles, while the evaporation rate can be adjusted through the incident laser power. Unlike sputtering, TLE does not require a process plasma, providing a comparatively simple environment for film growth. It also avoids the complexities associated with electron-beam evaporation such as contamination from the crucible. TLE has been demonstrated for nonradioactive solid elements across the periodic table \cite{Smart:2021} and has subsequently enabled the growth of epitaxial films of elemental and compound materials \cite{majer2024alpha, kim2021epitaxial, Kim:2025, hyatt2026growth}, including superconducting $\alpha$-Ta \cite{majer2024alpha} and TiN \cite{hyatt2026growth}. Epitaxial oxide and nitride films have been grown in oxygen/ozone mixtures and ammonia at pressures up to around $10^{-3}$ Torr.  \cite{kim2021epitaxial,hyatt2026growth}.  Recently, we reported nickel deposition using a home-built TLE system \cite{catherall2026characterization}. However, the properties of Nb films deposited by TLE have not been reported.

Here, we report the deposition and characterization of Nb films on c-plane sapphire substrates using TLE. The thin films were characterized by X-ray diffraction (XRD), atomic force microscopy (AFM), room-temperature van der Pauw (vdP) measurements, and temperature-dependent electrical transport. Two films deposited under nominally identical conditions exhibited reproducible properties; the film with the highest superconducting transition temperature exhibited $T_\mathrm{c} =  9.24$ K and a residual-resistivity ratio (RRR) of 5.71 with a thickness of $22.47 \pm 0.05$~nm. The present films exhibit near-bulk $T_\mathrm{c}$ and RRR values approximately 1.5--2.5 times those reported for Nb films of comparable thickness deposited using conventional techniques without substrate heating. This work indicates that TLE is a promising technique for deposition of superconducting refractory compounds.



\section*{Experimental Methods}

\subsection*{Film deposition}

Nb thin films were deposited on c-plane sapphire substrates (MTI Corporation, c-plane, $10 \times 10 \times 0.5$~mm, single-side polished) using the home-built TLE system described in our previous work \cite{catherall2026characterization}. Briefly, the system consists of a water-cooled ultrahigh-vacuum (UHV) chamber equipped with bellows-mounted target and substrate holders and a laser-delivery system, consisting of a fiber-laser coupler, collimator, and focusing optics. The laser is directed through a flange-mounted window onto the target using a water-cooled off-axis mirror. A continuous-wave 1070~nm fiber laser serves as the heating source. The target is press-fit into a molybdenum holder, with a type-C thermocouple positioned beneath it. The thermocouple serves primarily as a safety indicator to confirm the laser is hitting the target; its readings are not used to quantify the evaporation-surface temperature. The reflected laser beam is intercepted by a water-cooled beam dump.

In this work, a free-standing Nb target (Thermo Scientific, 99.95\%) was  heated by the laser to generate the evaporation flux. Several changes were made compared to our prior deposition of Ni. \cite{catherall2026characterization} First, the beam was focused more tightly to a diameter of approximately 500 \textmu m to achieve a higher power density. Second, we observed changes in the target surface morphology caused by prolonged localized laser heating at a fixed position, as has been reported previously \cite{Smart:2023_2}. These variations lead to flux instabilities. We addressed this problem by rastering the laser beam across the target surface, limiting localized deformation and helping to preserve the evaporation geometry and flux distribution during deposition. A similar strategy is used in pulsed-laser deposition to reduce localized target modification and promote uniform ablation \cite{fujioka2015pulsed}.

Another origin of flux instability is a change in mirror reflectivity caused by material deposition. An increase in the laser power required to maintain the deposition rate was initially observed, consistent with a decrease in mirror reflectivity caused by Nb deposition. This effect plateaued after several deposition runs, after which a reproducible growth rate could be achieved at a fixed laser power.

Before deposition, the chamber was evacuated to a base pressure of $2\times10^{-8}\,\mathrm{Torr}$, determined by an Edwards nWRG digital gauge. The Nb target and its holder were baked out at a laser power of 375~W. A residual gas analyzer (SRS RGA200) was used to evaluate the chamber vacuum condition. Desorbing water vapor was detected, and the deposition did not start until the partial pressure of the water dropped below $10^{-8}\,\mathrm{Torr}$. To reduce surface organic contamination, the as-received sapphire substrates were immersed in freshly prepared piranha solution ($\mathrm{H_2SO_4}:30\%~\mathrm{H_2O_2}=3:1$ by volume)
for 30~s and subsequently rinsed with deionized water. They were then mounted approximately 8~cm from the Nb target and held under vacuum overnight at room temperature before deposition. During deposition, the hydrogen partial pressure was observed to exceed $5\times10^{-7}$ Torr, consistent with the well-known affinity of Nb for hydrogen \cite{antoine2003h}. The substrates were not heated. No substrate rotation was employed. To begin deposition, the laser power was increased to 533~W and the substrate shutter was removed. The deposition was performed for 20 min. The film thickness was measured by X-ray reflectivity (XRR) performed with a step size of $0.002^\circ$ and a scan speed of $1^\circ$/min. The average deposition rate was obtained by dividing the measured thickness by the deposition time.

\subsection*{Surface and structural characterization}

Surface morphology was characterized by a Bruker Dimension Icon AFM in PeakForce tapping mode using a ScanAsyst-Air probe over scan areas of 500~nm $\times$ 500~nm. The images were processed using NanoScope Analysis, with first-order flattening applied to remove scanner background. The root-mean-square (RMS) surface roughness, \(R_\mathrm{q}\), was calculated over the scan area.

XRD measurements were performed using a Rigaku Smartlab diffractometer with a Cu-K$\alpha_1$ source. Symmetric $\theta$--$2\theta$ scans were collected with a step size of $0.02^\circ$ and a scan speed of $2.4^\circ$/min. Diffraction peaks were assigned by comparison with reference patterns for body-centered cubic (bcc) Nb and c-plane sapphire. The out-of-plane coherent domain size was estimated from the measured Nb peak width using the Scherrer equation without correction for instrumental broadening.

\subsection*{Electrical transport measurements}

Room-temperature sheet resistance was measured using the vdP method with four electrical contacts positioned near the sample perimeter by aluminum (Al) wire bonding using a TPT HB05 wire bonder. Measurements were performed in both reciprocal contact configurations with a Keithley 2400 SourceMeter using an excitation current of \SI{10}{\milli\ampere}. The film resistivity, $\rho$, was calculated as $\rho = R_{\mathrm{s}}t$,
where $R_{\mathrm{s}}$ is the measured sheet resistance and $t$ is the film thickness.

Temperature-dependent electrical resistivity was measured using a Physical Property Measurement System (PPMS, Quantum Design DynaCool) in a four-probe configuration with an excitation current of \SI{100}{\micro\ampere}. The sample was cooled from 300 to 20~K at 10~K/min, from 20 to 9.5~K at 1~K/min, from 9.5 to 8.5~K at 0.5~K/min, and from 8.5 to 2~K at 1~K/min. Measurements were acquired at nominal temperature intervals of 1, 0.05, 0.01, and 0.05~K over these respective ranges.


\section*{Results}

Two Nb films, denoted samples A and B, were deposited in separate runs under nominally identical conditions to assess the run-to-run reproducibility of the TLE system. The figures show the measured properties for sample A, while the corresponding properties of sample B are reported in the text for comparison.

\begin{figure}
    \centering{
    {\includegraphics[width=0.8\linewidth]{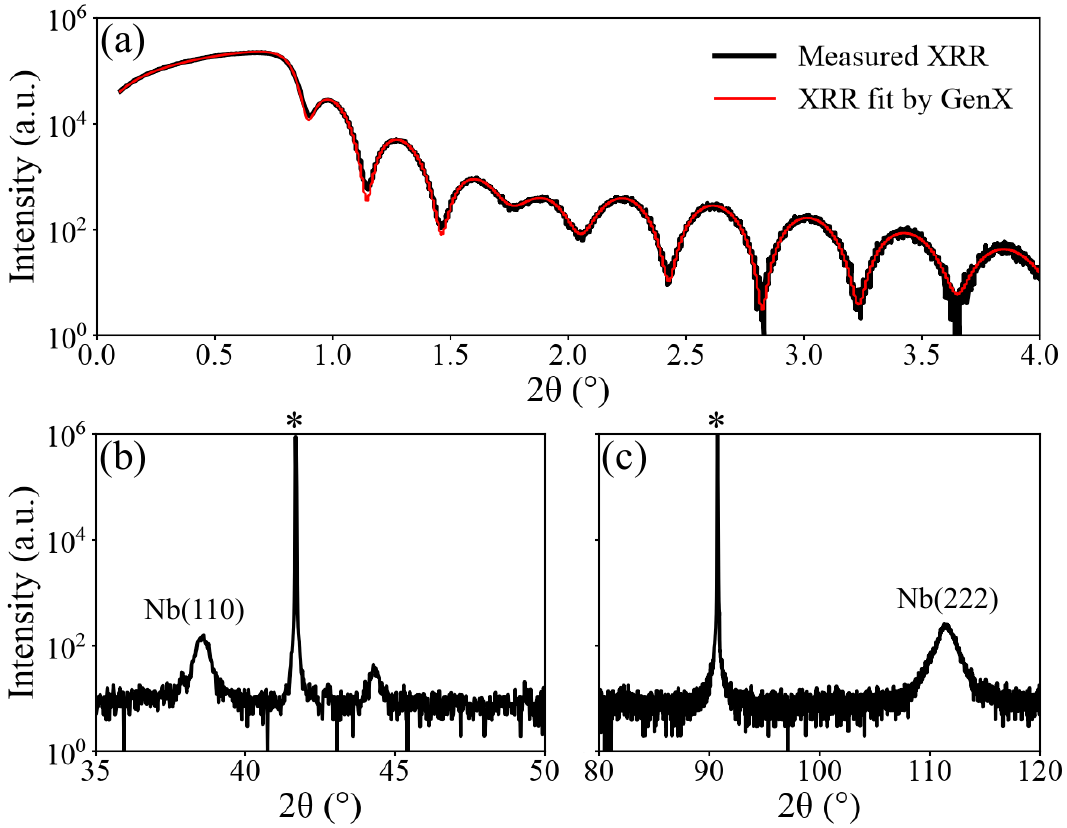}}
\phantomsubcaption\label{fig:xrr}
\phantomsubcaption\label{fig:xrd110}
\phantomsubcaption\label{fig:xrd222}
    }
    \caption{Structural characterization of sample A. (a) XRR versus angle (black line) and the corresponding GenX fit (red line), yielding a film thickness of $22.47 \pm 0.05$~nm. (b,c) Symmetric $\theta$--$2\theta$ XRD scans over different angular ranges using Cu-K$\alpha_1$ radiation. The diffraction peaks at $38.553 \pm 0.003^\circ$ and $111.494 \pm 0.005^\circ$ are assigned to Nb(110) and Nb(222), respectively. Asterisks (*) denote reflections from the c-plane sapphire substrate.}
    \label{fig:XRD}
\end{figure}

\Cref{fig:xrr} shows the XRR profile of sample A. The experimental data were fitted using GenX~\cite{glavic2022genx}, employing a multilayer model consisting of air/Nb$_2$O$_5$/Nb/sapphire. The fitted thicknesses of the Nb and Nb$_2$O$_5$ layers were $19.77 \pm 0.05$~nm and $2.70 \pm 0.05$~nm, respectively, yielding a total film thickness of $22.47 \pm 0.05$~nm. The thickness uncertainties were estimated from the parameter ranges corresponding to a 5\% increase in the minimum logarithmic figure of merit (FOM). Throughout this work, the term ``film thickness'' refers to the total thickness, including the oxide layer. The observed Kiessig fringes are consistent with well-defined interfaces and a relatively uniform film thickness. Based on the fitted total thickness and deposition time, the deposition rate was calculated to be approximately 0.19~\AA/s. The corresponding thicknesses of the Nb and Nb$_2$O$_5$ layers and the total film thickness of sample B were $20.63 \pm 0.11$~nm, $2.21 \pm 0.07$~nm, and $22.84 \pm 0.07$~nm, respectively, yielding a deposition rate of 0.19~\AA/s. The deposition rates obtained in this study are close to the rate reported for electron-beam evaporation of Nb for device fabrication \cite{kowsari2021fabrication, morohashi2001characteristics}. Previous TLE work reported a rate of approximately 0.34~\AA/s at 250~W (ref. \cite{Smart:2023_2}), whereas the present growth rate was lower even at a higher laser power, likely because the larger target size and press-fit mounting design of our system resulted in larger radiative and conductive heat losses. Higher deposition rates could be achieved by increasing the laser power or reducing the target size and its thermal coupling to the holder.

The $\theta$--$2\theta$ XRD pattern of sample A is shown in \Cref{fig:xrd110,fig:xrd222}. The diffraction peaks were fitted using a pseudo-Voigt line shape to extract the peak position $2\theta$ and the full width at half maximum of the diffraction peak $\beta$ (in radians). Their uncertainties represent one standard error, calculated from the diagonal elements of the covariance matrix of the least-squares fit. In addition to reflections from the c-plane sapphire substrate, peaks assigned to bcc Nb(110) and Nb(222) are observed at $2\theta =$~$38.553 \pm 0.003^\circ$ and $111.494 \pm 0.005^\circ$, respectively. Relative to the bulk reference positions of $38.545^\circ$ and $107.897^\circ$, these shifts correspond to out-of-plane compressive strains of $(-0.020 \pm 0.007)\%$ and $(-2.186 \pm 0.003)\%$, respectively, as calculated using Bragg's law. The strain uncertainties were propagated from the fitted peak-position uncertainties, treating the bulk reference positions as exact. The substantial compression inferred for the Nb[111]-oriented component remains unexplained, and further structural characterization is needed to confirm the origin of the peak shift. 

The out-of-plane coherent domain size was estimated using the Scherrer equation,
\begin{equation}
D=\frac{K\lambda}{\beta\cos\theta},
\end{equation}
where $D$ is the coherent domain size, $K$ is the Scherrer shape factor, and $\lambda$ is the X-ray wavelength. Using $K=0.9$ and $\lambda=1.5406~\text{\AA}$ for Cu-K$\alpha_1$ radiation, the resulting coherent domain sizes were $18.77\pm 0.40$~nm for Nb(110) and $8.70\pm 0.10$~nm for Nb(222). The uncertainties were propagated from the fitted $\beta$ and $\theta$, accounting for their covariance. These values describe the diffraction coherence length normal to the corresponding lattice planes. 

It is well established that substrate temperature can strongly influence thin-film texture and preferred crystallographic orientation, and Nb grown on c-plane sapphire is no exception. Drimmer \textit{et al.} observed a temperature-driven evolution from predominantly Nb[110]-oriented growth toward Nb[111]-oriented growth as the substrate temperature was increased to 975~K \cite{drimmer2025effect}. At even higher temperatures, Wagner reported the opposite orientation transition, with Nb\{111\} parallel to the sapphire basal plane at $900\,^{\circ}\mathrm{C}$ and Nb\{110\} at $1100\,^{\circ}\mathrm{C}$ \cite{wagner1998high}. These results highlight the strong dependence of Nb orientation on the specific growth conditions. However, the relatively strong Nb(222) reflection observed here is still unexpected based on previous studies \cite{sedai2026optimization, drimmer2025effect, chattaraj2025preventing}. A possible contributor to the observed Nb(222) reflection is the absence of substrate rotation during deposition. Without rotation, the local incidence angle of the incoming vapor flux varies across the substrate and is not azimuthally averaged during growth. This spatially varying deposition geometry may alter the local growth kinetics and, consequently, the preferred crystallographic orientation across the film. 

An additional diffraction peak was observed at $44.343 \pm 0.008^\circ$, which could not be assigned to a Bragg reflection of bcc Nb. A separate XRD scan of the bare aluminum sample plate confirmed that this peak originates from the sample plate. Sample B also exhibits the same Nb(110) and Nb(222) reflections, with corresponding Bragg peaks at $38.512 \pm 0.010^\circ$ and $110.972 \pm 0.006^\circ$, and coherent domain sizes of $19.99 \pm 1.55$~nm and $5.37 \pm 0.04$~nm, respectively. Relative to the same bulk reference positions, these peak shifts correspond to an out-of-plane tensile strain of $(0.082 \pm 0.025)\%$ for the Nb[110]-oriented component and a compressive strain of $(-1.881 \pm 0.004)\%$ for the Nb[111]-oriented component.


\begin{figure}
    \centering{
    \includegraphics[width=0.6\linewidth]{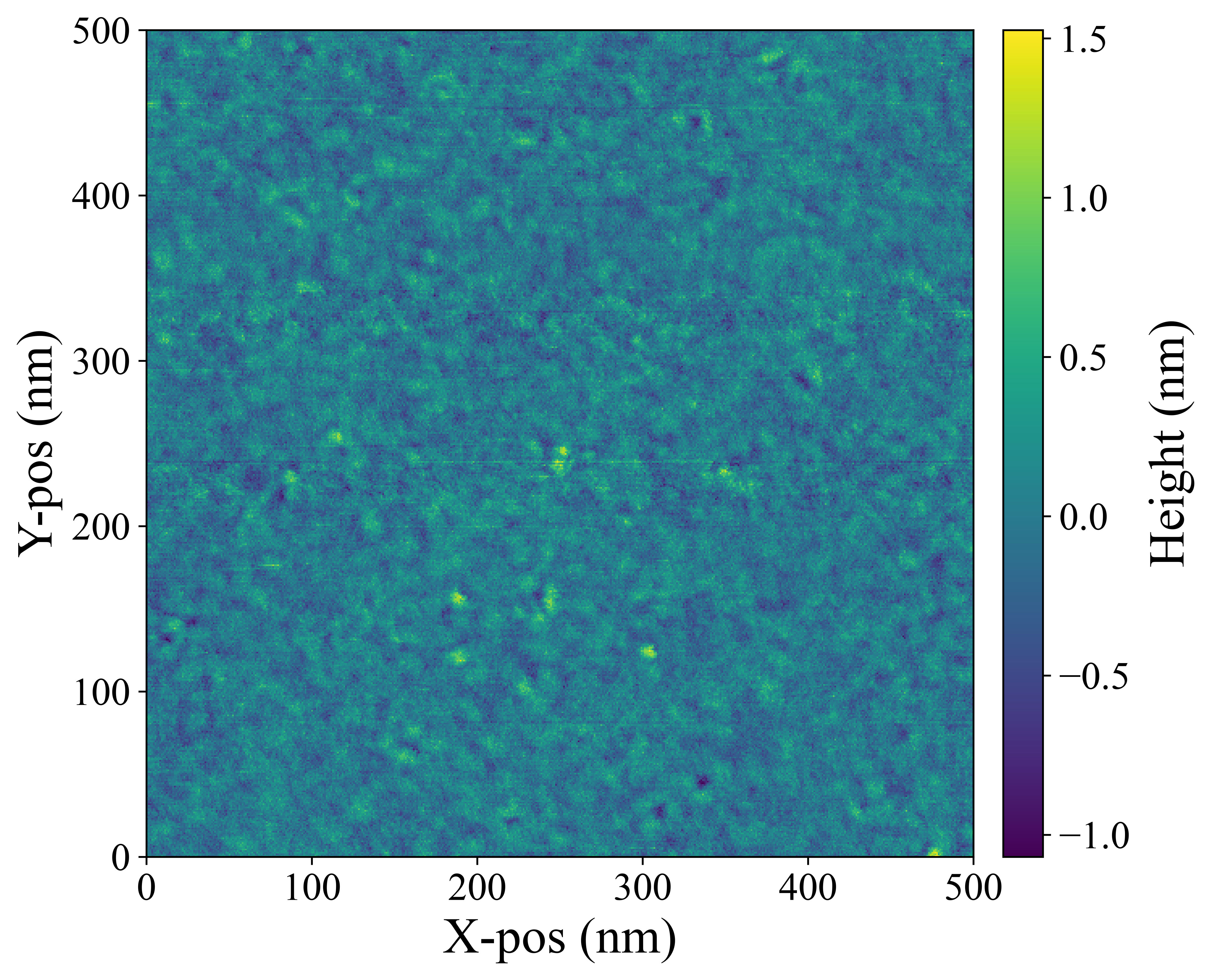}}
    \caption{AFM topography of sample A over a 500~nm $\times$ 500~nm scan area. The RMS surface roughness is $R_q = 0.23 \pm 0.02$~nm, where the uncertainty represents the standard deviation of the roughness values calculated from individual line scans. The image was processed using NanoScope Analysis with first-order flattening applied. The color scale represents surface height in nanometers.
}
    \label{fig:AFM}
\end{figure}

\Cref{fig:AFM} shows an AFM topographic image of sample A over a 500~nm $\times$ 500~nm area. The film exhibits a smooth and uniform surface within the scanned region, with an RMS roughness, $R_\mathrm{q}=0.23 \pm 0.02$~nm. Sample B has a comparable $R_\mathrm{q} = 0.28 \pm 0.03$~nm. The uncertainty represents the standard deviation of the roughness values calculated from individual line scans. This low surface roughness is consistent with the pronounced Kiessig fringes observed by XRR, whose visibility indicates smooth surface and buried interfaces, together with good thickness uniformity across the X-ray-illuminated area. The AFM and XRR data therefore provide complementary evidence for the smoothness and uniformity of the film.

\begin{figure}
    \centering{
    {\includegraphics[width=1.0\linewidth]{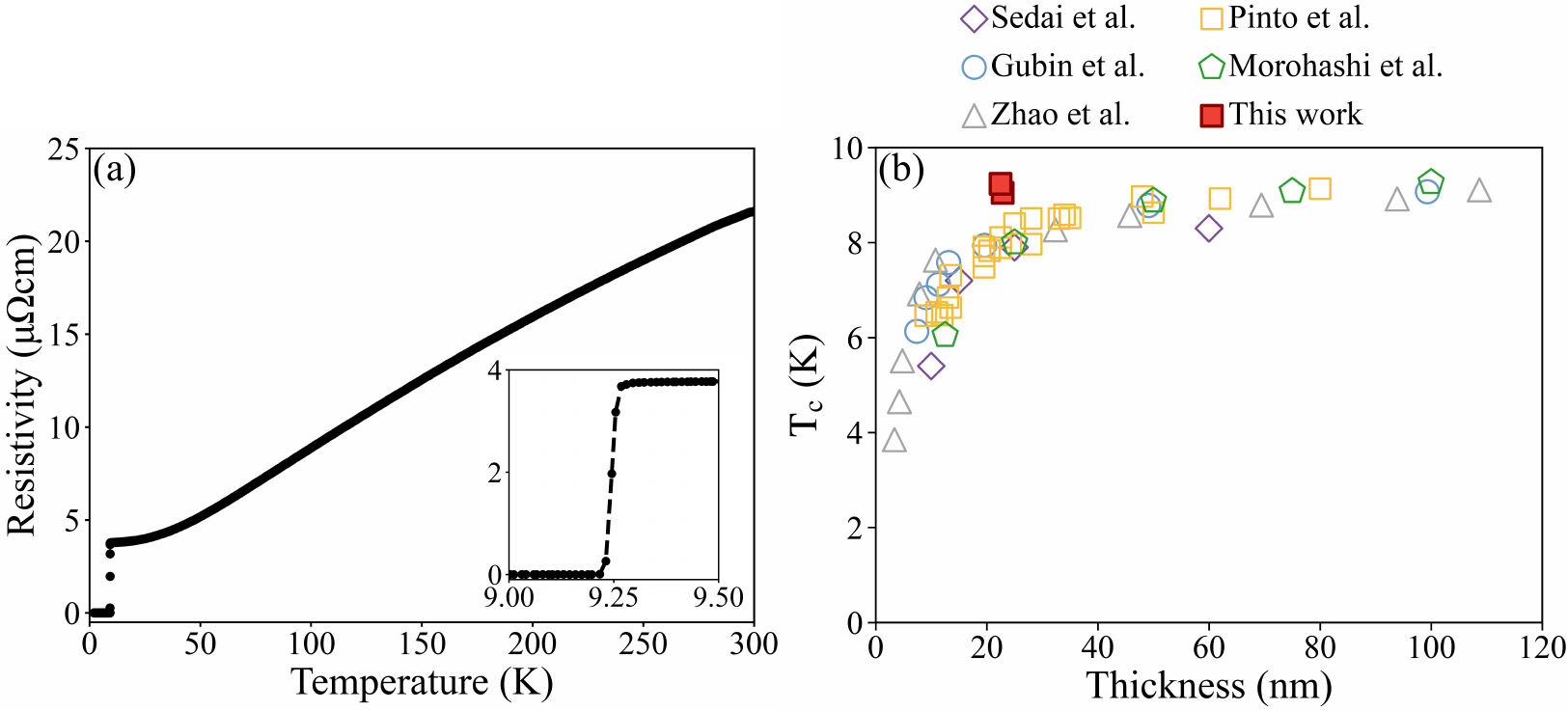}}
    \phantomsubcaption\label{fig:tc_a}
    \phantomsubcaption\label{fig:tc_b}}
    \par\vspace{-20pt}
    \caption{(a) Measured electrical resistivity versus temperature. The inset shows a magnified view of the superconducting transition region. The dashed line is a guide to the eye. (b) Superconducting transition temperature of Nb thin films versus thickness from this work (filled red squares) and other techniques including DC magnetron sputtering in  Ref.~\onlinecite{sedai2026optimization} (open purple diamonds), Ref.~\onlinecite{gubin2005dependence} (open blue circles), Ref.~\onlinecite{zhao2014fabrication} (open grey triangles), and Ref.~\onlinecite{pinto2018dimensional} (open yellow squares), and e-beam evaporation in Ref.~\onlinecite{morohashi2001characteristics} (open green pentagons).}
    \label{fig:Tc}
\end{figure}

\begin{figure}
    \centering{
    {\includegraphics[width=1.0\linewidth]{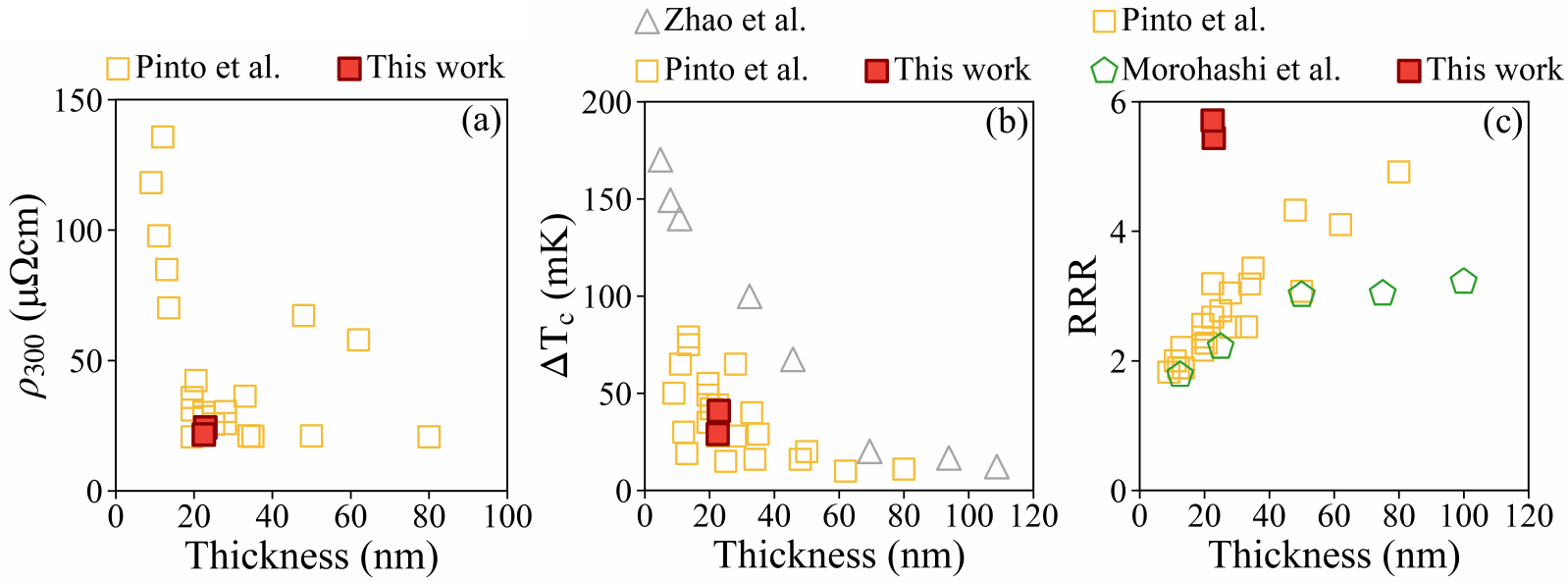}}
    \phantomsubcaption\label{fig:rho}
    \phantomsubcaption\label{fig:delta_tc}
    \phantomsubcaption\label{fig:RRR}}
    \par\vspace{-20pt}
    \caption{(a) Room-temperature resistivity, (b) superconducting transition width, and (c) RRR versus thickness from this work (filled red squares) and previous studies. Data from Ref.~\onlinecite{pinto2018dimensional} (open yellow squares), Ref.~\onlinecite{zhao2014fabrication} (open grey triangles), and Ref.~\onlinecite{morohashi2001characteristics} (open green pentagons) are also shown for comparison.}
    \label{fig:Comparison}
\end{figure}

\Cref{fig:tc_a} shows the temperature-dependent resistivity of sample A. The raw resistance data were calibrated with respect to the room-temperature resistivity, $\rho_\mathrm{300}=21.60 \pm 0.05$~\textmu$\Omega \mathrm{cm}$, calculated from the sheet resistance measured by the standard vdP method and the thickness characterized by XRR. The reported resistivity uncertainty was propagated from the XRR thickness uncertainty, neglecting uncertainty in the measured sheet resistance. For reference, the bulk resistivity of Nb is 15.2~\textmu$\Omega \mathrm{cm}$ at 273~K \cite{lide2003crc}, and thin films typically exhibit higher resistivity due to grain boundary and surface scattering. 

The normal-state resistivity, $\rho_\mathrm{n}$, was taken as the resistivity at 10~K. The transition temperature was defined by the midpoint criterion, $\rho(T_\mathrm{c})=0.5\rho_\mathrm{n}$, yielding $T_\mathrm{c}=9.24$~K. The transition width was quantified as $\Delta T_\mathrm{c}=T_{90}-T_{10}=29$~mK,
where $T_{90}$ and $T_{10}$ are the temperatures at which the resistance reaches 90\% and 10\% of the normal-state value, respectively. RRR, defined as $\rho_\mathrm{300}/\rho_\mathrm{n}$, is 5.71. Sample B exhibits a room-temperature resistivity of $\rho_\mathrm{300}=24.32 \pm 0.07$~\textmu$\Omega \mathrm{cm}$, a transition temperature of 9.05~K, a transition width of 41~mK and an RRR of 5.44, showing comparable transport properties between the two deposition runs.

\Cref{fig:tc_b} compares the superconducting transition temperatures of samples A and B with values reported for Nb films deposited by direct-current (DC) magnetron sputtering~\cite{sedai2026optimization,gubin2005dependence,zhao2014fabrication,pinto2018dimensional} and electron-beam evaporation~\cite{morohashi2001characteristics}. Both TLE-grown films exhibit $T_\mathrm{c}$ values close to the bulk value of Nb and lie above the reported data at comparable thicknesses. In particular, whereas the reported transition temperature of conventionally deposited Nb films generally decreases appreciably as the thickness approaches the 20--30~nm range, the TLE-grown films retain $T_\mathrm{c}>9$~K at a thickness of approximately 22.5~nm. This demonstrates that TLE preserves near-bulk superconducting transition temperatures at lower film thickness compared to prior reports.

To further assess the electrical quality of the TLE-grown films, \cref{fig:Comparison} compares the room-temperature resistivity, superconducting transition width, and residual-resistivity ratio with literature data. As shown in \Cref{fig:rho}, the measured room-temperature resistivities of the TLE films, $\approx 21.60$ and $24.32$~\textmu$\Omega \mathrm{cm}$, fall within the lower range of values reported for sputtered Nb films of comparable thickness. Pinto \textit{et al.} reported $\rho_{300}=25.6$~\textmu$\Omega \mathrm{cm}$ for a 25-nm-thick Nb film, while substantial sample-to-sample variations ranging from approximately 20 to 42~$\mu\Omega \mathrm{cm}$ were observed for films in the 20--30~nm thickness range~\cite{pinto2018dimensional}. The room-temperature resistivity of the TLE films is therefore comparable to that of conventionally deposited Nb films.

The superconducting transition widths are compared in \Cref{fig:delta_tc}. Samples A and B, with a thickness of around 22.5 nm, exhibit transition widths of 29 and 41~mK, respectively. These values are comparable to those reported by Pinto \textit{et al.} \cite{pinto2018dimensional} and substantially narrower than the values reported by Zhao \textit{et al.} \cite{zhao2014fabrication} in a similar thickness range. Pinto \textit{et al.} \cite{pinto2018dimensional} reported $\Delta T_\mathrm{c}$ values of approximately 15--65~mK for films between 20 and 30~nm, while Zhao \textit{et al.} \cite{zhao2014fabrication} observed a rapid increase in the transition width up to approximately 170~mK as the film thickness decreases below 10~nm. Thus, the transition widths of the TLE-deposited samples fall well within the range of characteristics typical of high-quality Nb films. The narrow transition widths further suggest that the near-bulk $T_\mathrm{c}$ values are accompanied by spatially uniform superconducting properties \cite{tinkham1996introduction}. A small $\Delta T_\mathrm{c}$ indicates a narrow distribution of local $T_\mathrm{c}$ values over the film and is commonly associated with high-quality Nb films \cite{pinto2018dimensional}.

\begin{figure}
    \centering{
    {\includegraphics[width=1.0\linewidth]{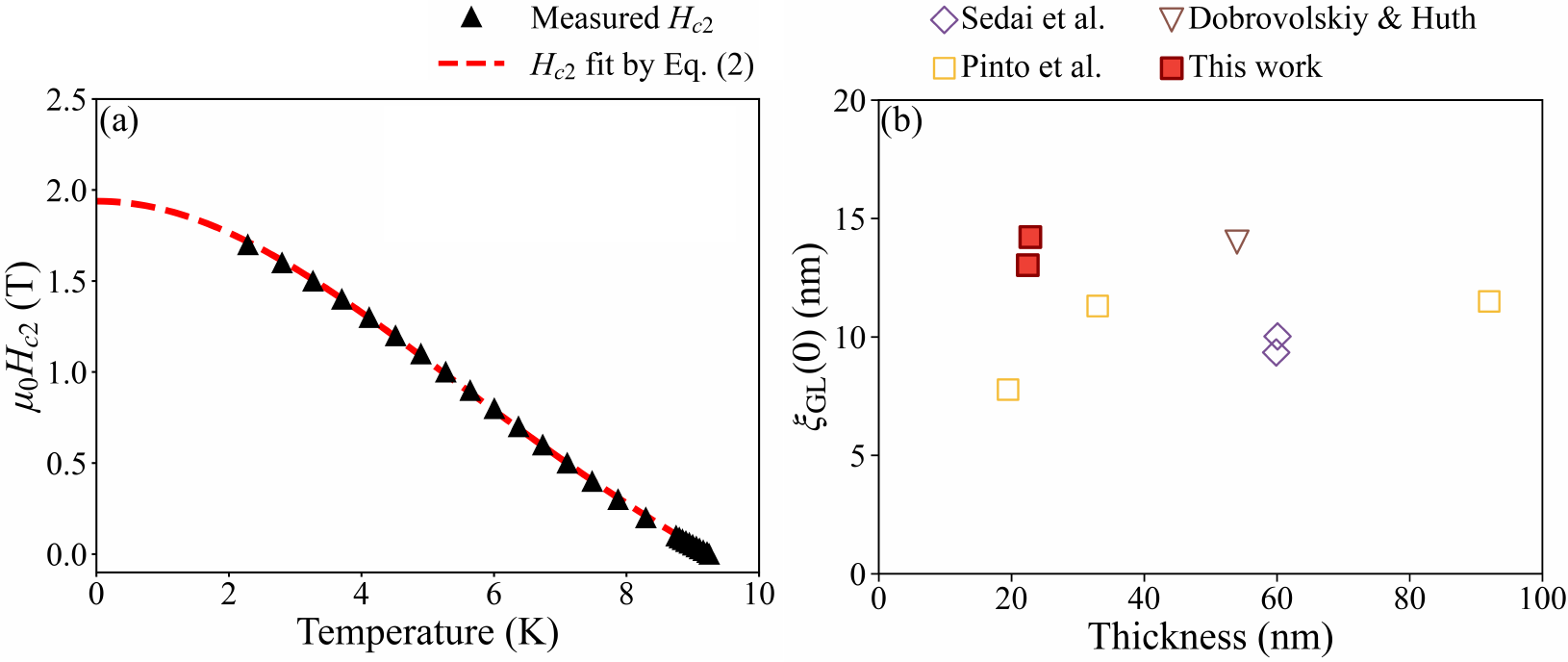}}
    \phantomsubcaption\label{fig:Hc2_a}
    \phantomsubcaption\label{fig:Hc2_b}}
    \par\vspace{-20pt}
    \caption{(a) Upper critical field of sample A as a function of temperature. Red dashed curve shows the best fit by Eq. (2). (b) Coherence length of Nb thin films at different thicknesses. Present work (filled red squares); Ref.~\onlinecite{sedai2026optimization} (open purple diamonds); Ref.~\onlinecite{pinto2018dimensional} (open yellow squares); Ref.~\onlinecite{dobrovolskiy2012crossover} (open brown downward triangles).}
    \label{fig:Hc2}
\end{figure}

A more pronounced distinction is observed in the residual-resistivity ratio, as shown in \Cref{fig:RRR}. The two TLE-grown films exhibit RRR values of 5.71 and 5.44. These values substantially exceed those of the sputtered Nb films reported by Pinto \textit{et al.} \cite{pinto2018dimensional} at comparable thicknesses, for which RRR is typically between approximately 2 and 3.5. The values also exceed those reported for high-vacuum electron-beam-evaporated Nb by Morohashi \textit{et al.} \cite{morohashi2001characteristics}, where RRR approaches about 3.2 at a thickness of 100~nm. Since RRR is sensitive to residual defect and impurity scattering, the combination of high RRR and the room-temperature resistivity comparable to that of reported Nb films deposited conventionally suggests a low degree of normal-state disorder in the TLE-grown Nb films.

The temperature dependence of the upper critical field, $\mu_0 H_\mathrm{c2}(T)$, of sample A is shown in \Cref{fig:Hc2_a}. The upper critical field was determined from resistivity measurements under magnetic fields applied perpendicular to the film plane. For each field, the sample was first stabilized at 12~K, above zero-field $T_\mathrm{c}$, and the magnetic field was applied. The sample was then cooled through the full superconducting transition and subsequently heated back to 12~K before the field was changed to the next value. From each resulting resistive transition, $T_\mathrm{c}$ was identified using the same 50\% resistance criterion employed for the zero-field transition, despite the progressive broadening of the transition with increasing magnetic field. The resulting $\mu_0 H_\mathrm{c2}(T)$ values were fitted using the empirical expression~\cite{karki2011physical,pinto2018dimensional}
\begin{equation}
\mu_0 H_\mathrm{c2}(T)
=
\mu_0 H_\mathrm{c2}(0)
\frac{1-(T/T_\mathrm{c})^2}
{1+(T/T_\mathrm{c})^2},
\label{eq:Hc2_ratio}
\end{equation}
yielding $\mu_0 H_\mathrm{c2}(0)=1.938 \pm 0.004$~T, where the uncertainty represents one standard error estimated from the covariance matrix of the least-squares fit. Using the relation between zero-temperature Ginzburg--Landau (GL) coherence length and the upper critical field,
\begin{equation}
\mu_0 H_{\mathrm{c2}}(0)
=
\frac{\Phi_0}{2\pi \xi_{\mathrm{GL}}^2(0)},
\label{eq:Hc2_xi}
\end{equation}
where $\Phi_0 = 2.07 \times 10^{-15}\,\mathrm{Wb}$ is the magnetic flux quantum, the coherence length was calculated to be $\xi_\mathrm{GL}(0)=13.031 \pm 0.012$~nm. The uncertainty was propagated from the fitted upper critical field. The same measurement and analysis were applied to sample B, yielding $\mu_0 H_\mathrm{c2}(0)=1.627 \pm 0.005$~T and $\xi_\mathrm{GL}(0)=14.224 \pm 0.024$~nm. Note that owing to the experimental schedule, the critical-field measurements on sample B were performed approximately two months after deposition. Although the sample was stored in a glovebox during this period, its $T_\mathrm{c}$ decreased to 8.78~K at the time of measurement. A possible explanation for this reduction is gradual surface oxidation during storage, which likely reduced the effective film thickness.

\Cref{fig:Hc2_b} compares the coherence lengths of the TLE-grown samples with values reported for Nb thin films in the literature~\cite{pinto2018dimensional,sedai2026optimization,dobrovolskiy2012crossover}. Because the reported studies employ different criteria for determining $H_\mathrm{c2}(T)$ and different models for extrapolating $H_\mathrm{c2}(0)$, this comparison should be regarded as qualitative. Nevertheless, the coherence lengths of the TLE-grown Nb films fall within the range reported for high-quality Nb thin films with thicknesses below 100~nm, providing an additional indication that the superconducting properties of the TLE-grown films are comparable to those obtained using established deposition techniques.

\section*{Discussion}

The Nb films deposited by TLE exhibit near-bulk $T_\mathrm{c}$ and RRR values approximately 1.5--2.5 times those reported for Nb films of comparable thickness deposited using conventional techniques without substrate heating. Together with the measured surface roughness and other transport and structural properties, these results provide a quantitative picture of the characteristics of Nb thin films deposited by TLE. They complement previous demonstrations of epitaxial superconducting Ta and TiN using thermal laser epitaxy \cite{majer2024alpha, hyatt2026growth}. 

Building on this work, several technical upgrades to our TLE system would expand the accessible growth conditions and enable synthesis of refractory compounds. Introducing substrate rotation could improve deposition uniformity across the substrate and provide an additional parameter for controlling film texture. Substrate heating, for instance using a CO$_2$-laser substrate-heating system that is installed on our system, would enable in situ high-temperature substrate degassing immediately before growth, reducing the reliance on prolonged room-temperature degassing under vacuum and providing better control over substrate preparation. More importantly, active substrate heating would extend TLE beyond unheated deposition and enable systematic studies of temperature-dependent nucleation, texture, and epitaxial growth phenomena. The integration of an N$_2$ plasma source would further introduce a controllable source of reactive nitrogen, extending the system toward reactive growth of refractory nitrides and carbonitrides. Together, these upgrades would make the system a more versatile platform for investigating the growth of refractory metals and other high-temperature materials.

As a potential direction for future work, the addition of  substrate heating could be used to further explore the superconducting properties of TLE-grown Nb. In the supplemental measurements of Glotzer \textit{et al.}, elemental Nb grown on sapphire at $600^{\circ}\mathrm{C}$ by electron-beam evaporation exhibits a superconducting transition near 9.6~K, exceeding the bulk value of approximately 9.25~K \cite{glotzer2026thermally}. Given that the Nb films demonstrated here already reach a near-bulk $T_\mathrm{c}$ without substrate heating, the introduction of substrate temperature as an additional growth parameter may provide a route toward further optimization of the Nb film structure and potentially allow superconducting transition temperatures to exceed those obtained in the present unheated growth conditions.

Beyond Nb and superconducting materials, the combination of controlled substrate heating and a reactive N$_2$ plasma source would also open access to a broader class of refractory compounds. A future direction for TLE is the controlled synthesis of binary and ternary transition-metal carbides and nitrides, particularly Zr- and Hf-based systems, to investigate how surface thermodynamics, growth kinetics, impurities, and vapor fluxes govern thin-film phase formation and microstructure. Although these materials have been studied extensively in bulk, uncertainties remain in their phase equilibria and thermophysical properties, while thin-film growth introduces additional phenomena through surface energies, kinetic limitations, and strain \cite{ushakov2019carbides}. TLE could therefore provide a useful platform for systematically exploring both equilibrium and metastable phases under well-controlled conditions compared to sputtering or other PVD techniques. This direction is particularly interesting for the Hf--C--N system, where first-principles calculations predict carbonitride compositions with the highest melting temperatures among all known materials \cite{hong2015prediction}.  


\section*{Conclusion}

In summary, superconducting Nb thin films were deposited on c-plane sapphire by TLE. The films exhibited smooth surfaces, well-defined crystalline textures, and reproducible superconducting properties, with the highest $T_\mathrm{c}$ of 9.24~K obtained for a $22.47 \pm 0.05$~nm-thick film. The TLE-grown films exhibit near-bulk superconducting transition temperatures and high RRR values relative to previous reports. These results indicate that TLE can deposit high-quality superconducting Nb thin films and motivate further studies of a broader range of deposition conditions and refractory compounds.

\section*{Acknowledgements}
The authors thank Marionne Epalle for useful discussions and Jeong Rae Kim for assistance with the XRD measurements. The TLE system was acquired under Air Force Office of Scientific Research (AFOSR) Award FA9550-23-1-0731 and an award from the De Logi Foundation at Caltech. D.S.C. and A.J.M. were supported by Office of Naval Research (ONR) Award N000142612339. Y.Y. was supported by an Explorer Grant from the Resnick Sustainability Institute at Caltech. F.B.D. acknowledges support from the National Science Foundation Graduate Research Fellowship Program. This material is based upon work supported by the National Science Foundation Graduate Research Fellowship Program under Grant No. 2139433. Any opinions, findings, and conclusions or recommendations expressed in this material are those of the author(s) and do not necessarily reflect the views of the National Science Foundation. We gratefully acknowledge the critical support and infrastructure provided for this work by the Kavli Nanoscience Institute at the California Institute of Technology for the use of their facilities.

\section*{Data Availability Statement}
The data that support the findings of this study are available from the corresponding author upon reasonable request.

\section*{Conflict of Interest}
The authors have no conflicts to disclose.

\bibliographystyle{is-unsrt}
\bibliography{bib}

\end{document}